# π-Plasmon Dispersion in Free-Standing Graphene by Momentum-Resolved Electron Energy-Loss Spectroscopy

S. C. Liou,[1,2] C.-S. Shie,[3] C. H. Chen,[1] R. Breitwieser,[1] W. W. Pai,[1] G. Y. Guo,[3*] and M.-W. Chu[1,4*]

[1]*Center for Condensed Matter Sciences, National Taiwan University, Taipei 10617, Taiwan*

[2]*Maryland NanoCenter, University of Maryland, College Park, MD 20742*

[3]*Department of Physics, National Taiwan University, Taipei 10617, Taiwan*

[4]*Taiwan Consortium of Emergent Crystalline Materials, Minister of Science and Technology, Taipei 10617, Taiwan*

## ABSTRACT

The π-plasmon dispersion in graphene was scrutinized by momentum($q$)-resolved electron energy-loss spectroscopy with an improved $q$ resolution and found to display the square root of $q$ dispersion characteristic of the collective excitation of two-dimensional electron systems, in contrast with previous experimental and theoretical studies which reported a linear $q$ dispersion. Our theoretical elaborations on the $q$-dependent spectra affirm this square root of $q$ relation and further unveil an in-plane electronic anisotropy. The physical property of the π plasmon is thoroughly compared to that of the two-dimensional plasmon due to carriers of the Dirac fermions. A clear distinction between the π plasmon and the two-dimensional Dirac plasmon was demonstrated, clarifying the common notion on correlating the linearly-dispersed Dirac cones to the linear dispersion of the π plasmon previously reported.

## I. INTRODUCTION

Graphene, a single layer of carbon atoms densely packed in a honeycomb lattice, is the first stable two-dimensional (2D) material found in ambiance.[1,2] This 2D-carbon sheet comprises a conical-band character of Dirac fermions at the corners of Brillouin zone (BZ) showing a distinct linear $q$ dependence,[1-3] and graphene continues to be the model system in the search for emergent 2D phenomena, ranging from topological states to anomalies in monolayer transition-metal dichalcogenides.[4,5]

In graphene, the 2D Dirac fermions are massless and exhibit remarkably high, robust carrier mobility upon temperature variations.[1-3] These features add a plethora of novel device applications,[1] in analogy of 2D electron gas (2DEG) in conventional semiconductor heterojunctions. Using $q$-dependent electron energy-loss spectroscopy (EELS), the collective plasmon excitation of the 2D electrons (with density generally in the order of ~$10^{13}$ cm$^{-2}$) is found to form an otherwise dispersive feature at 0 ~ 1 eV, with the characteristic $\sqrt{q}$-dependence distinctly different from the linearly-dispersed Dirac cones of a single-particle essence.[3,6-12] This $\sqrt{q}$-dispersion has been theoretically investigated in detail and demonstrated to be the signature of collective quasiparticle excitations with a 2D nature.[3,6-9] Such a 2D Dirac plasmon with its low excitation energy boasts another intriguing opportunity of graphene, plasmonics applications in the near(far)-infrared regime.[10,11] Indeed, the technological merit of graphene continues to grow with the increasing understanding of the Dirac fermions.[1,2,12] Apart from these intriguing characteristics, the elementary structure of graphene is nonetheless determined by the $\pi$ and $\sigma$ valence electrons lying outside the Dirac cones, with the former being unpaired in an out-of-plane orbital and the latter being hybridized with the three coordinated carbons in $ab$-plane, as parent graphite.[1-3,13,14]

In graphene and also parent graphite, the $\pi$ and $\sigma$ electrons manifest the related collective excitations of $\pi$ and $\pi+\sigma$ plasmons above ~4 and ~15 eV, respectively.[15-22] Moreover, both the collective excitations of parent graphite show a quadratic dispersion (proportional to $q^2$) in line with the typical dynamical response in 3D bulks.[3,20-22] By contrast, graphene is a perfectly 2D matter and its collective $\pi$ and $\pi+\sigma$ plasmons oscillate in the 2D plane, intuitively giving rise to a 2D character of the plasmons just like the 2D Dirac plasmon at 0 ~ 1 eV.[3,6-9] Surprisingly, three recent $q$-dependent EELS studies of the $\pi$ plasmon of graphene reported a quasi-linear dispersion,[17-19] which is a general characteristic of the plasmon excitations of one-dimensional electron systems,[3,16] and a separate theoretical work on the subject also pointed out the same dispersion feature.[23] These works represent the detailed works on the $\pi$-plasmon dispersions of graphene,[17-19,23] while the linear dispersion derived is at odds with the established notion on the $\sqrt{q}$-dispersion characteristic of 2D systems.[3,6-9] An elaborate $q$-dependent EELS investigation, which largely escapes experimental scrutiny due to the central attention on the 2D Dirac plasmon,[3,6-11] is essential for shedding light on this inconsistency.

In this work, we report $q$-dependent EELS studies of the $\pi$ plasmon in free-standing graphene with an improved $q$ resolution of ~0.001 Å$^{-1}$ over the entire BZ and a momentum transfer up to the zone boundary of ~1.5 Å$^{-1}$. The previous $q$-dependent EELS studies showed a $q$ resolution in the order of ~0.01 Å$^{-1}$ and/or accessed only a fraction of BZ.[6-8,16-19] With the fine $q$ resolution and extensive $q$ range, we were able to explore the $\pi$ plasmon along the two principal in-plane directions, $\overline{\Gamma M}$ and $\overline{\Gamma K}$, at high precision and consistently found the $\sqrt{q}$-dispersion characteristic of 2D-collective excitations. A further tilting experiment of the graphene, allowing out-of-plane contributions to the EELS excitation, indicated that the $\pi$-plasmon dispersion is free from any out-of-plane dispersive component, unveiling the true 2D character of the $\pi$ plasmon without ambiguity.[3,4,6-9] The EELS results along with an in-plane electronic anisotropy observed at large $q$ were theoretically investigated and the dispersion of the $\pi+\sigma$ plasmon was also discussed.

## II. EXPERIMENTS AND THEORETICAL CALCULATIONS

The EELS experiments were performed on a transmission electron microscope (TEM, FEI Tecnai F20), operated at 120 kV and equipped with a field-emission gun. To achieve the superb $q$ resolution required for resolving the $\pi$-plasmon dispersion of graphene unambiguously, an exceptionally long camera length of the diffraction pattern is crucial and this can be attained by raising the sample-object plane so as to use the graphene as an optical grating of the incident parallel illumination by itself (detailed methodology in Ref. 24), resulting in a camera length of ~92 m. Further considering the EELS-slit size of 4×10$^{-4}$ m and the finite diffraction spot focused on the slit, a $q$-resolution of ~0.001 Å$^{-1}$ was derived. The accompanied EELS energy resolution is ~0.6 eV.

The graphene sheet was grown on a Cu foil by chemical vapor deposition and then transferred onto a TEM grid in a free-standing form. A careful monitoring of the diffraction-spot intensity revealed the monolayer feature in many patches (few tens of micrometers in the lateral dimension) of the thus-prepared sample,[25] and the EELS results were acquired from these regions. The EELS experiments of graphite were conducted on exfoliated natural graphite with a thickness of ~40 nm.

The theoretical understanding of our EELS observations was undertaken in the framework of the density functional theory with the local density approximation (LDA) plus the linear response of random phase approximation (RPA) and otherwise adiabatic LDA (ALDA).[26,27] We use the accurate real-space projector augmented wave function (PAW),[28] which is implemented in the GPAW code.[29] Graphene is simulated by a slab-supercell with the in-plane lattice constant ($a$) of 2.46 Å and the elevated interlayer spacing of 25 Å in order to assure a good convergence. A grid spacing of 0.2 Å was used throughout all calculations. In addition, a $k$-sampling with $25 \times 25 \times 1$ for the 2D-BZ was exploited in the self-consistent calculation of the band structure. A denser $k$-point mesh of $100 \times 100 \times 1$ and a plane-wave energy cutoff up to 200 eV (i.e., including 819 plane waves) were taken into account upon the evaluation of the theoretical $q$-dependent EELS spectra, which further integrate the 50 bands up to 40 eV above the Fermi level. For comparison, we have also performed the same calculations for graphite. The lattice constants of $a$ = 2.46 Å and $c$ = 6.71 Å were used. The $k$-sampling mesh of $20 \times 20 \times 7$ was considered in the self-consistent band structure calculation and a fine $k$-point mesh of $40 \times 40 \times 14$ was exploited in the calculation of the $q$-dependent dielectric function and EELS spectra.

### III. RESULTS AND DISCUSSION

The $q$-resolved EELS experiments of graphene were performed along both $\overline{\Gamma M}$ and $\overline{\Gamma K}$. In Fig. 1a along $\overline{\Gamma M}$, the superb $q$ resolution facilitates a direct observation of the dispersive $\pi$ and $\pi+\sigma$ plasmons in the long-wavelength limit, starting from ~4 and ~13 eV at $q \to 0$, respectively. The dispersive feature of these excitations in such a small $q$-range (0 ~ 0.012 Å$^{-1}$) has never been resolved due to a compromised $q$-resolution of the previous reports (0.03 ~ 0.06 Å$^{-1}$).[17-19] Although the strong elastic peak tends to saturate the intensity of Fig. 1a below ~3 eV, both plasmons point to a non-linear dispersion, which will be affirmed shortly in Figs. 2-3.

In Fig. 1b, we show the EELS spectra extracted from Fig. 1a with $q \approx 0.002 \sim 0.012$ Å$^{-1}$. The integrated spectrum over the whole $q$ range is also exhibited (bottom gray curve, Fig. 1b) and the thus-indicated $\pi$- and $\pi+\sigma$-plasmon peaks at ~4.5 and ~15 eV (see also the inset), respectively, are consistent with the well-known excitation energies resolved by scanning TEM (STEM) that typically integrates over $q$ as a result of the convergent-beam optics.[15] This agreement with the STEM results reaffirms the benefit and necessity of performing EELS experiments with high $q$ resolution under this circumstance and, with this reinforced confidence level of our $q$-resolved setup, we now proceed to the dispersion over the whole BZ along $\overline{\Gamma M}$ and $\overline{\Gamma K}$ (Fig. 2).

Figs. 2a and 2b show the large-$q$ dispersion along $\overline{\Gamma M}$ up to the BZ boundary (~1.5 Å$^{-1}$) and the EELS spectra acquired at selected $q$'s, respectively. Fig. 2c exhibits the corresponding large-$q$ dispersion along $\overline{\Gamma K}$. We note from Fig. 2 that the $\pi$ plasmon disperses from ~4 ($q \to 0$) to ~12 eV ($q \sim 1.5$ Å$^{-1}$) along both $\overline{\Gamma M}$ and $\overline{\Gamma K}$ with an accompanied broadening of the peak as expected. The dispersion of the $\pi+\sigma$ plasmon from ~13 ($q \to 0$) to ~30 eV ($q \sim 1.5$ Å$^{-1}$) can also be resolved. Intriguingly, the $\pi$-plasmon dispersion along $\overline{\Gamma M}$ is accompanied with a low-energy, dispersive shoulder for $q$ larger than 0.5 Å$^{-1}$, while such a phenomenon is absent along $\overline{\Gamma K}$. We will come back to this feature and also the dispersion of the $\pi+\sigma$ plasmon later, and we now focus on the $\pi$-plasmon dispersion outlined in Fig. 3.

In Fig. 3a, the experimental dispersions of the $\pi$ plasmon along ΓQ and ΓP of parent graphite (counterparts of respective $\overline{\Gamma M}$ and $\overline{\Gamma K}$) are also shown and both exhibit the characteristic parabolic dependence ($\propto q^2$).[20-22] In addition, the dispersions along the two in-plane directions are indistinguishable in the $q$ range of $0 \sim 0.5$ Å$^{-1}$ and start to deviate from each other at $q$ above 0.5 Å$^{-1}$. In graphene (Fig. 3a), it is obvious that, toward the long-wavelength limit ($0 \sim 0.5$ Å$^{-1}$), the $\pi$-plasmon dispersions along $\overline{\Gamma M}$ and $\overline{\Gamma K}$ are neither quadratic as the bulk graphite nor linear as previously reported,[17-19,23] although the dispersions tend to mimic those of parent graphite for $q$ larger than 0.5 Å$^{-1}$. A non-linear feature of the $\pi$-plasmon dispersions of graphene is indeed resolved in the $q$ range of $0 \sim 0.5$ Å$^{-1}$ (Fig. 3a) and the associated dispersions along $\overline{\Gamma M}$ and $\overline{\Gamma K}$ are almost the same. Now, we rescale the $\pi$-plasmon excitation energy as a function of $\sqrt{q}$ in Fig. 3b. Notably, the non-linear $\pi$-plasmon dispersion with $q$ smaller than 0.5 Å$^{-1}$ is faithfully underlined by this $\sqrt{q}$

-dependence (Fig. 3b), indicative of plasmons with a 2D character.[3,6-9] In principle, a perfectly 2D excitation is to be further supported by the absence of an out-of-plane electronic components of the state.[4] We carefully examined this possibility by tilting the graphene specimen (Fig. 4), the technique of which has been exploited to reveal the anisotropic in-plane and out-of-plane signatures of the $\pi$- and $\pi+\sigma$-plasmon in 3D parent graphite.[20,30]

At the first glance on Fig. 4a, the $\pi$-plasmon dispersion curves acquired upon three separate tilting angles ($\alpha = 0, 45,$ and $60°$) are visibly different, showing a decrease in excitation energies with increasing $\alpha$ and, therefore, seemingly suggesting the existence of an out-of-plane factor like parent graphite. Nevertheless, it should be noted from the inelastic scattering kinematics (inset, Fig. 4a) that, upon the tilting, the effective momentum transfer is the $q$ component projected onto the graphene sample, $q_s$ ($\approx q\cos\alpha$), rather than the primitive $q$. A rescaling of Fig. 4a as a function of $q_s$, shown in Fig. 4b, reveals not only the profound equivalence of the three dispersion curves, but, more importantly, the absence of any out-of-plane dispersive contribution to the $\pi$ plasmon.[4] The 2D character of the $\pi$ plasmon of graphene is now established. Although the $\pi+\sigma$ plasmon is too broad for such a sample-tilting inspection, a detailed examination of Fig. 3c indicates that the $\pi+\sigma$-plasmon dispersion basically scales with the $\sqrt{q}$ -relation as the $\pi$ plasmon and also reflects its 2D character.

With all these EELS elaborations in Figs. 2-4, both the $\pi$ and $\pi+\sigma$ plasmons of graphene are clearly of a 2D essence. The previous reports of a linearly-dispersed $\pi$ plasmon should be a consequence of the limited $q$-resolution therein.[17-19] It is also noted that a recent theoretical report on the collective excitations in graphene suggests a quadratic dispersion of the $\pi+\sigma$ plasmon,[23] which is not found at all in our EELS investigations in Fig. 3c. Moreover, in Fig. 2 we also did not observe a splitting of the $\pi+\sigma$ plasmon with a magnitude of few eV for $q$ larger than ~1.0 Å$^{-1}$ only along $\overline{\Gamma K}$ as reported in a recent EELS study of the plasmon dispersions of graphene (Ref. 17). A splitting of this magnitude should be easily resolvable with our energy resolution of ~0.6 eV if it exists. This $\pi+\sigma$-plasmon splitting issue would require future investigations.

Having established the 2D character of the $\pi$ and $\pi+\sigma$ plasmons of graphene, we turn to tackle the dispersive low-energy shoulder accompanied with the $\pi$ plasmon, showing up only along $\overline{\Gamma M}$ for $q$ larger than 0.5 Å$^{-1}$ and apparently further broadening the π-plasmon peaks (Figs. 2b and 3a). Indeed, the same EELS feature was also reported in Ref. 17 and qualitatively suggested as a π-plasmon splitting, however, without addressing its physical origin. In the relevant theoretical work of Ref. 23, a similar splitting along $\overline{\Gamma M}$ was found to disperse throughout the *q* range unlike the experimental observations in Figs. 2b and 3a and Ref. 17. Therein, the dispersive shoulder was understood as the π-plasmon splitting without further details.[23] The splitting of a surface plasmon, which commonly occurs due to the electromagnetic coupling between two adjacent surfaces, is not expected to happen in single layer graphene.[31] Therefore, we attempted to address the electronic origin of this low-energy shoulder along $\overline{\Gamma M}$, and a close examination of Figs. 3a-b reveals three interesting characteristics. First, the same dispersive shoulder appears in parent graphite along ΓQ for *q* larger than ~0.5 Å$^{-1}$ (Fig. 3a, open diamond). Second, the π-plasmon dispersions of graphite and graphene are basically identical in the *q*-regime above 0.5 Å$^{-1}$ (Fig. 3a), with the π-plasmon dispersion along ΓQ ($\overline{\Gamma M}$) lying above that along ΓP ($\overline{\Gamma K}$) in graphite (graphene). Third, the scaling of the π-plasmon dispersions along $\overline{\Gamma M}$ and $\overline{\Gamma K}$ deviates from the $\sqrt{q}$-relation when *q* becomes larger than 0.5 Å$^{-1}$ (Fig. 3b).

In parent graphite, the low-energy dispersive shoulder arises from a direct, nonvertical π → π* interband transition present along ΓQ at large *q* and plays the role of blue-shifting the associated π plasmon accordingly.[20-22] This thus-shifted π plasmon then sits on top of the dispersion curve along ΓP as observed in Fig. 3a, indicating an in-plane electronic anisotropy originated from the different characteristic band structures along the two inequivalent ΓQ and ΓP directions.[20-22] Indeed, the two in-plane counterpart directions of graphene, $\overline{\Gamma M}$ and $\overline{\Gamma K}$, are also intrinsically inequivalent. If ignoring the linearly-dispersed Dirac cones at the $\overline{K}$ point, the band structures of graphene and graphite are otherwise similar.[2,13,32] For instance, the π → π* interband transition of graphene also occurs at ~4 eV at the $\overline{M}$ point,[2,13,15,32] and the corresponding band dispersion closely resembles that of graphite.[32] The low-energy dispersive feature along $\overline{\Gamma M}$ could then be associated with a direct, nonvertical interband transition like graphite, as confirmed by our *ab-initio* calculations of the *q*-dependent electronic excitations in graphene and graphite within both RPA and ALDA (Fig. 5).

Figs. 5a-b show the calculated *q*-dependent EELS spectra of graphite along the respective ΓQ and ΓP, and Figs. 5c-d exhibit the readily derived π-plasmon dispersions and the low-energy dispersive feature. The experimental EELS results of graphite in Fig. 3a are also incorporated into Figs. 5c-d for comparison. Indeed, both the parabolic π-plasmon dispersion of graphite and the direct, nonvertical π → π* transition along ΓQ at large $q$ are nicely captured in our calculations (Figs. 5c-d). In Figs. 5e-f, the graphene-counterpart calculations are also notably consistent with the corresponding EELS observations of the $\sqrt{q}$-scaling of the π-plasmon dispersion and the onset of the low-energy dispersive transition at $q \sim 0.5$ Å$^{-1}$. For $q < 0.5$ Å$^{-1}$ where the π-plasmon excitation is predominant, the electronic screening effect is strong due to a large real part of the complex dielectric function ($\varepsilon = \varepsilon_1 + i\varepsilon_2$) as shown in the corresponding calculations in Fig. 6a and the intensity of the low-energy interband transition features is readily overwhelmed by the plasmon (see the associated loss function), becoming invisible in EELS. For $q > 0.5$ Å$^{-1}$, the π-plasmon oscillation is, however, increasingly damped with a diminishing intensity and the lower-energy peak due to the interband transition can then emerge as a shoulder (see the calculated loss function, Fig. 6b).

Nonetheless, there exists a systematic overestimation for the ALDA results compared to the RPA calculations of graphene (Figs. 5e-f). Indeed, the ALDA method is optimal for 3D matters with a nearly homogeneous electron density and the previous ALDA calculations of such a material, Al, have been found to satisfactorily depict the characteristic plasmon dispersion.[33] By contrast, graphene is purely 2D and the associated electron density changes abruptly along the out-of-plane direction, thus not an ideal geometry for ALDA. For the ALDA method to be more appropriate for 2D systems, an improved exchange-correlation potential would be indispensable and represents an intriguing challenge to be resolved in the future. Otherwise, the general agreement between the theoretical calculations and experiments in Fig. 5 points to a close electronic similarity between 2D graphene and 3D graphite for $q$ larger than 0.5 Å$^{-1}$ as a result of the intimate resemblance of the electronic structures therein between them.[32] This electronic similarity underlines the three former features of Figs. 3a-b as a whole and, more importantly, establishes the in-plane electronic anisotropy of graphene just like the parent graphite.

The electronic excitations of graphene above ~4 eV are investigated in detail by far. We intend to further compare the π plasmon to the 2D Dirac plasmon at 0 ~ 1 eV.[3,7,8]

In the previous π-plasmon studies of graphene,[17,18,23] the reported linear dispersion, proven otherwise incorrect herein, has been correlated with transitions from the linearly-dispersed Dirac cones[18] and became a generally accepted notion.[17,23] The possibility for the Dirac fermions to entangle with the π plasmon is vanishingly small, since the spectral weight of carriers in the Dirac cones centers below ~1 eV due to its small carrier density (~$10^{13}$ cm$^{-2}$),[7,8,10-12] which is two orders of magnitude below that of π-valence electrons as discussed later in the following. The correlation of the reported linear dispersion of the π plasmon to the linear Dirac cones can be safely discounted. Nonetheless, the π-plasmon onset at ~4 eV as $q \to 0$ (Fig. 3a) coincides with the $\pi \to \pi^*$ vertical interband transition of ~4 eV at the $\overline{\mathrm{M}}$ point.[15,32] This raises the possibility of mixed collective and single-particle essences for the π plasmon,[19] and an electromagnetic oscillation of this type, so-called plexciton (coupling of interband transitions and plasmon oscillators), has been reported.[8,34]

The continuous π-plasmon dispersion from ~4 ($q \to 0$) to ~12 eV ($q \sim 1.5$ Å$^{-1}$) and its $\sqrt{q}$-scaling in Fig. 3b follows the equation, $\omega_p(q) = \beta + \sqrt{\gamma q}$, derived for a 2D-collective excitation due to an interband transition with $\omega_P$ being the dispersive plasmon energy, $\beta$ the single-particle oscillator strength, and $\gamma = \sqrt{2\pi n_{2D} e^2 / m\varepsilon}$.[6,9] It should be noted that the derivation of $\gamma$ has been based on the long-range Coulomb response of non-interacting 2DEG to an external longitudinal electric field,[6,9] analogous to the longitudinal excitations in our EELS,[31] where $n_{2D}$ is the 2D electron density in cm$^{-2}$, $e$ is the elementary charge, $m$ is the effective mass of electron (0.06~0.07 $m_0$; $m_0$, the rest mass of electron),[13,14] and $\varepsilon$ is the dielectric constant (unity, for convenience). The square root of $\gamma$ was determined to be ~5.18 from Fig. 3b and the dispersion relation of π plasmon can now be written as $\omega_P \approx 4 + 5.18\sqrt{q}$ for $q$ smaller than 0.5 Å$^{-1}$. We were then able to obtain the corresponding 2D π-electron density as ~$2 \times 10^{15}$ cm$^{-2}$, notably in the same order as the total π electrons integrated in the first BZ of graphene, ~$3.8 \times 10^{15}$ cm$^{-2}$ ($\approx 4/\sqrt{3}a^2$).[14] A difference in the π-electron densities by a factor of two can be noted, implying that, under the consideration of the non-interacting 2DEG model,[6,9] only about half of the total π electrons participate in the corresponding 2D-plasmon excitation with the rest of the oscillator strength being taken up by the single-particle interband transition nearly at the same energy as the π plasmon. Hence, the agreement between the two evaluated electron densities can be a reasonable one, and the consistency between the thus-deduced single-particle oscillator strength (*β*) of ~4 eV and the associated interband transition (~4 eV) is satisfactory, altogether suggesting a single-particle mixture to the collective 2D *π* plasmon. Indeed, such a suggestion was also raised recently on the basis of the proposal that the linear dispersion of the *π* plasmon observed could be entangled with a linearly-dispersed *π* → *π** interband transition at lower energy along $\overline{\Gamma M}$ at $q < 0.5$ Å$^{-1}$.[19] Although we have firmly established the $\sqrt{q}$-dispersion of the *π* plasmon at $q < 0.5$ Å$^{-1}$ and the onset of the *π* → *π** transition only above 0.5 Å$^{-1}$ for $q$ along $\overline{\Gamma M}$, this coincidence in the suggestion is still intriguing. Nevertheless, we are conservative about further terming the *π* plasmon of graphene as plexciton, which features a coherent coupling between the single-particle and collective oscillator strengths.[34] The broad *π* plasmon observed herein (Figs. 2b-c) does not seem to support this element of coherent coupling.

## IV. CONCLUSION

Using *q*-dependent EELS, we have unveiled the 2D essence of the *π* and *π*+*σ* plasmons of graphene by the convincing observation of a characteristic $\sqrt{q}$-dispersion of 2D collective excitations toward the long-wavelength limit. For the *π* plasmon, evidence for the absence of any dispersive component along the out-of-plane direction, which is the further signature of a 2D excitation, was also presented. Besides, a low-energy, dispersive shoulder accompanied with the *π*-plasmon dispersion along $\overline{\Gamma M}$, but not along the other principal vector of $\overline{\Gamma K}$ in *ab*-plane, was theoretically tackled and found to arise from a direct, non-vertical interband transition along $\overline{\Gamma M}$ only, revealing an in-plane electronic anisotropy. The quantitative evaluation of the $\sqrt{q}$-scaling of the *π* plasmon microscopically reveals that this 2D collective oscillation is also electronically intermingled with the single-particle *π* → *π** interband transition, totally different from the well-known 2D Dirac plasmon of graphene with a purely Dirac-fermions contribution.

## ACKNOWLEDGEMENT

This work was supported by the Minister of Science and Technology, National Taiwan University, and Academia Sinica.

**FIGURE CAPTIONS**

**FIG. 1.** (Color online) (a) The $q$-dependent EELS map in the long-wavelength limit of graphene along $\overline{\Gamma M}$. Gray line, the light line. Curved dashed lines, guides for the eyes for the characteristic non-linear dispersions derived from the peak positions revealed in (b). (b) The EELS spectra extracted from (a) at the indicated $q$'s. Bottom gray line, the EELS spectrum integrated over the whole $q$ range of $0 \sim 0.012$ Å$^{-1}$ in (a). Inset, the blow-up of the $\pi+\sigma$-plasmon portion of the gray spectrum.

**FIG. 2.** (a) The $q$-dependent EELS map toward the BZ boundary of graphene along $\overline{\Gamma M}$. (b) The EELS spectra acquired at the selected $q$'s. (c) The $q$-dependent EELS spectra toward the BZ boundary along $\overline{\Gamma K}$. The spectra in (b) and (c) were all normalized to the spectral intensity at ~4.5 eV and then displaced vertically to improve the readability.

**FIG. 3.** (Color online) (a) The $\pi$-plasmon dispersions in graphene along respective $\overline{\Gamma M}$ (black square) and $\overline{\Gamma K}$ (red circle), and those in parent graphite along the $\Gamma Q$ and $\Gamma P$ counterparts as respective gray open diamonds and red open circles. (b) and (c), The rescaling of the $\pi$- and $\pi+\sigma$-plasmon dispersions along $\overline{\Gamma M}$ (black square) and $\overline{\Gamma K}$ (red circle) as a function of the $\sqrt{q}$-relation. Dashed gray line, guide for the eyes for the linear $\sqrt{q}$-scaling toward the long-wavelength limit in (b) and throughout the $q$ range in (c).

**FIG. 4.** (Color online) (a) The $\pi$-plasmon dispersions along $\overline{\Gamma M}$ as a function of the tilting angle ($\alpha$) of the graphene specimen from the normal incidence, with $\alpha = 0$ (black square), 45° (red circle), and 60° (gray triangle). Inset, the inelastic scattering kinematics at a given sample tilting of $\alpha$. $k_0$, incident beam; $k_f$, inelastically scattered beam; $q$ (blue), the momentum transfer corresponding to the inelastic scattering; $q_s$ (red, $\approx q\cos\alpha$), the exact momentum transfer on the graphene sheet. (b) The rescaled dispersion curves as a function of $q_s$.

**FIG. 5.** (Color online) (a) and (b), The calculated $q$-dependent EELS spectra of parent graphite along respective $\Gamma Q$ and $\Gamma P$ within the *ab-initio* RPA (black) and ALDA (red)

frameworks. (c) and (d), The dispersion curves derived from (a) and (b) along respective ΓQ and ΓP. The experimental EELS results in Fig. 3a are also shown for comparison (black square). (e) and (f), The calculated π-plasmon dispersions in graphene along respective $\overline{\Gamma M}$ and $\overline{\Gamma K}$ within the *ab-initio* RPA (red open circle) and ALDA (blue open triangle) methods and also the associated EELS experiments in Fig. 3a (black square).

**FIG. 6.** (Color online) The *q*-dependent complex dielectric function, $\varepsilon = \varepsilon_1 + i\varepsilon_2$, and the thus-derived electron-energy loss function, Im(-1/$\varepsilon$), of graphene along $\overline{\Gamma M}$ for (a) $q = 0.295$ Å$^{-1}$ and (b) $q = 1.003$ Å$^{-1}$ from the RPA calculations.

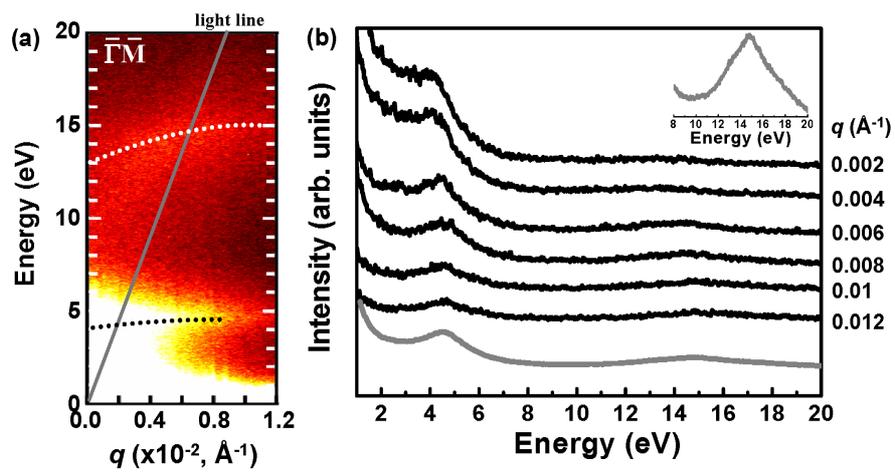

FIG. 1.

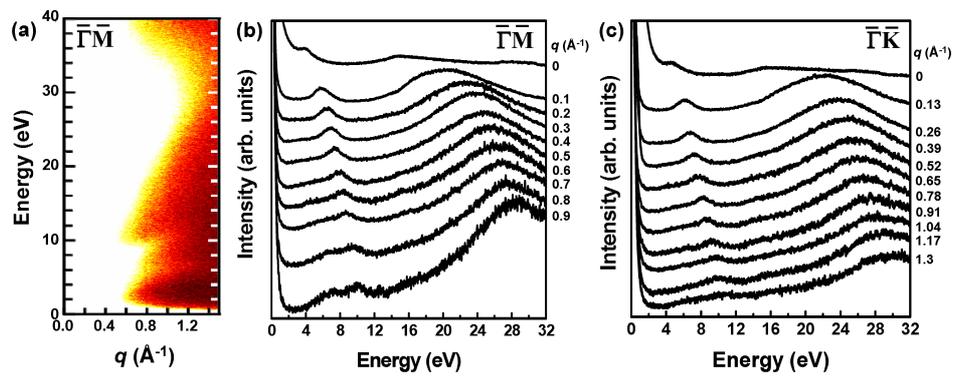

**FIG. 2.**

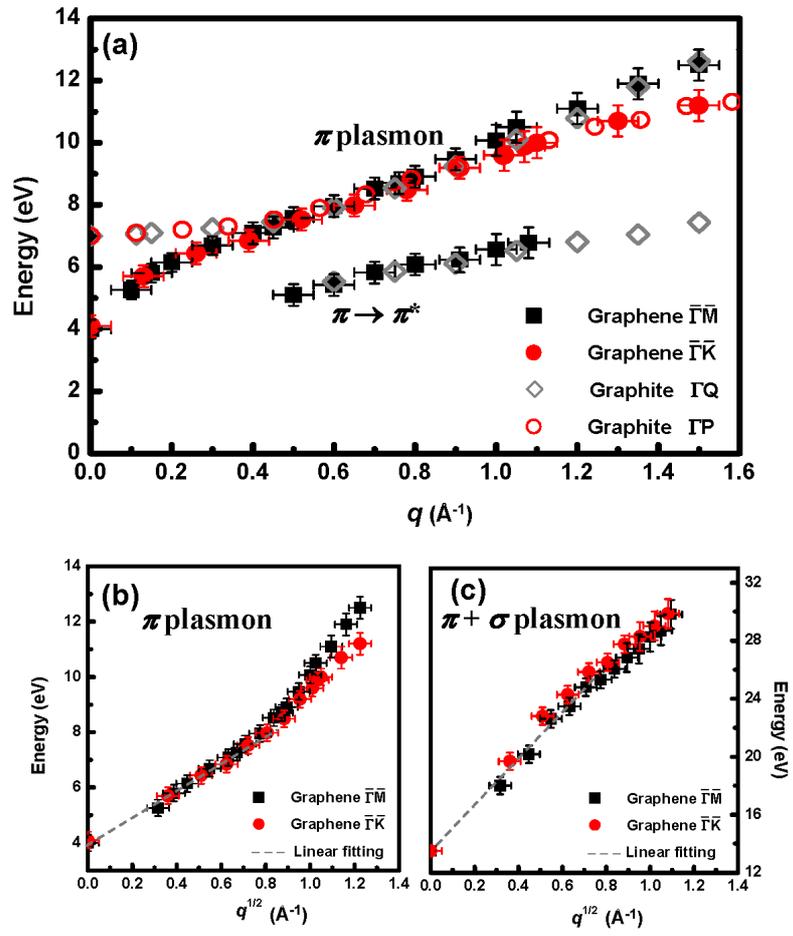

**FIG. 3.**

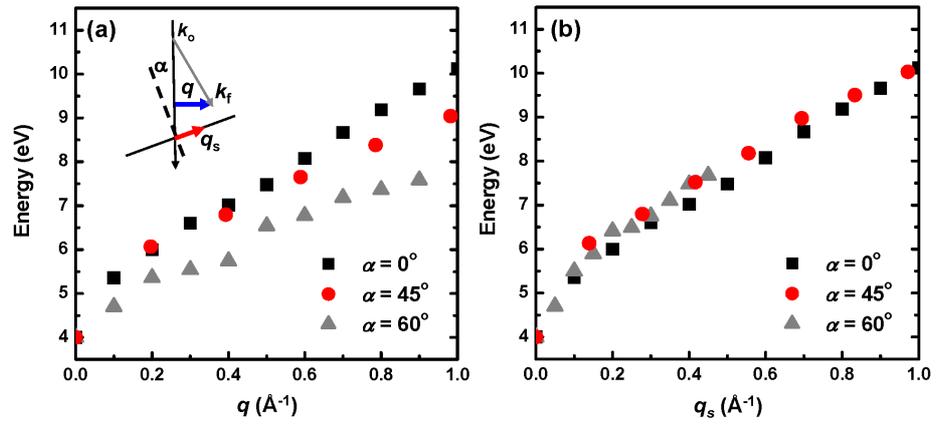

**FIG. 4.**

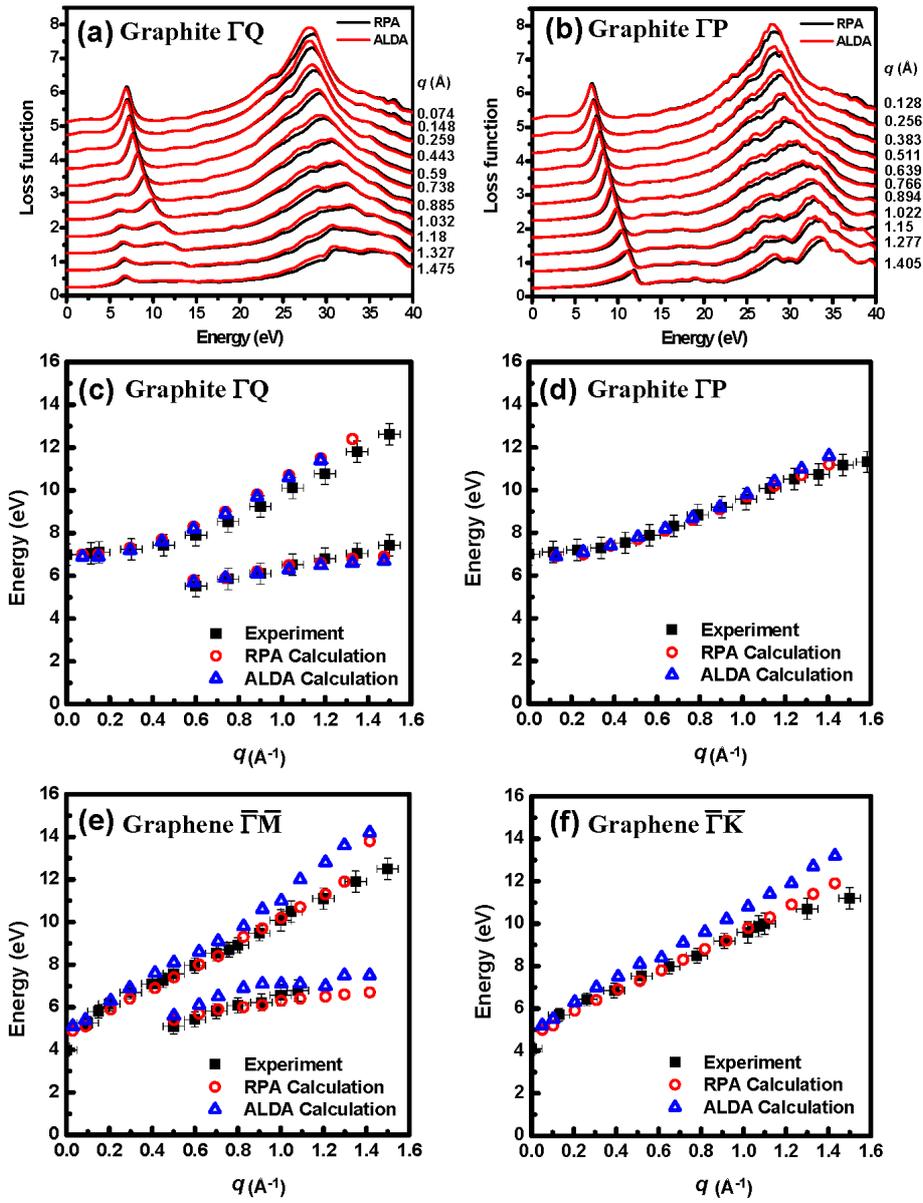

**FIG. 5.**

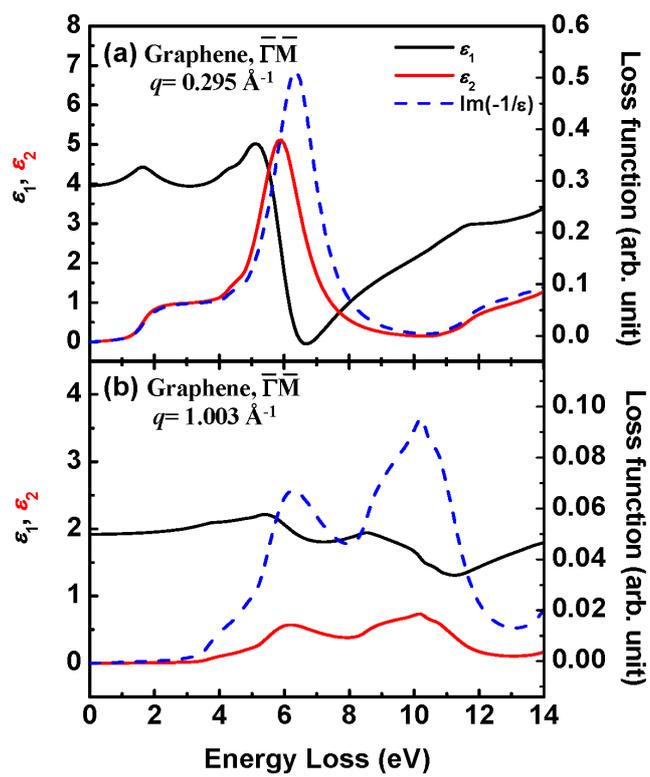

FIG. 6.